# A realistic dimension-independent approach for charged defect calculations in semiconductors

Jin Xiao[1,2,#], Kaike Yang[1,#], Dan Guo[1,3], Tao Shen[1,3], Jun-Wei Luo[1,3*], Shu-Shen Li[1,3], Su-Huai Wei[4*], and Hui-Xiong Deng[1,3*]

[1]*State Key Laboratory of Superlattices and Microstructures, Institute of Semiconductors, Chinese Academy of Sciences, Beijing 100083, China*

[2]*School of Science, Hunan University of Technology, Zhuzhou 412007, China*

[3]*Center of Materials Science and Optoelectronics Engineering, University of Chinese Academy of Sciences, Beijing 100049, China*

[4]*Beijing Computational Science Research Center, Beijing 100193, China*

[#] *These two authors contributed equally.*

[*] *Corresponding authors. E-mail address: jwluo@semi.ac.cn, suhuaiwei@csrc.ac.cn, and hxdeng@semi.ac.cn*

**First-principles calculations of charged defects have become a cornerstone of research in semiconductors and insulators by providing insights into their fundamental physical properties. But current standard approach using the so-called "jellium model" has encountered both conceptual ambiguity and computational difficulty, especially for low-dimensional semiconducting materials. In this Communication, we propose a physical, straightforward, and dimension-independent universal model to calculate the formation energies of charged defects in both three-dimensional (3D) bulk and low-dimensional semiconductors. Within this model, the ionized electrons or holes are placed on the realistic host band-edge states instead of the virtual jellium state, therefore, rendering it not only naturally keeps the supercell charge neutral, but also has clear physical meaning. This realistic model reproduces the same accuracy as the traditional jellium model for most of the 3D semiconducting materials, and**

**remarkably, for the low-dimensional structures, it is able to cure the divergence caused by the artificial long-range electrostatic energy introduced in the jellium model, and hence gives meaningful formation energies of defects in charged state and transition energy levels of the corresponding defects. Our realistic method, therefore, will have significant impact for the study of defect physics in all low-dimensional systems including quantum dots, nanowires, surfaces, interfaces, and 2D materials.**

## 1 Introduction

Doping semiconductors by introducing defects or impurity atoms is fundamental to controlling the properties of semiconductors and is the basis of all functionality in modern electronic and optoelectronic devices[1-5]. Physically, the doping efficiency of a semiconductor is characterized by the defect transition energy level, which can be defined as the Fermi energy $\varepsilon_F$ at which the formation energy $\Delta H_f(q)$ of the dopant in charge state $q$ is equal to the formation energy $\Delta H_f(q')$ in charge state $q'$[6-8]. Over the last three decades, first-principles defect calculations have made tremendous advances in accurately predicting defect transition energies and formation energies of bulk semiconductors, thus providing key insights in the fundamental processes of defect formation and charge carrier generation that are not otherwise easily accessible through experiment[8,9]. In modern first-principles density functional theory (DFT) defect calculations, a supercell approach is often used in which the defect is put at the center of the supercell of the host materials and a periodic boundary condition is applied[10]. As long as the supercell size is sufficiently large to shield the coupling of defects with their periodic images, this approach can accurately predict the formation energy of neutral defect[11]. For charged defects, $q$ electrons are removed (added) from (to) the defects and added (removed) to (from) the host states with Fermi energy level $\varepsilon_F$ and the charge neutrality condition is satisfied[12]. For example, for a donor defect $D$ with its donor transition energy level $E_D$ referenced to the host conduction band minimum (CBM) with energy $\varepsilon_C$, at T=0, the electron can be removed from the defect and placed at the host CBM with $\varepsilon_F=\varepsilon_C$. At finite temperature, the removed electron can occupy more states with energy levels $\varepsilon_i$ above the CBM with the occupation probabilities determine by the Fermi-Dirac distribution $f(\varepsilon_i) = \frac{1}{1+\exp[(\varepsilon_i-\varepsilon_F)/kT]}$ for a given $\varepsilon_F$ (Fig. 1a). In most practical calculations of charged defects in bulk semiconductors, the charge distribution of the removed or added electrons in the host band edge states are approximated by a virtual "jellium" charge (i.e., uniform charge distribution over the whole supercell) with energy level equals $\varepsilon_F$. This jellium model approximation is justified for most semiconductors because their host band edge states are indeed delocalized, yielding a similar charge distribution as that in the jellium model (Fig. 1b)[13-16].

However, the failure of this standard jellium model has been reported in two-dimensional (2D) semiconductors[17-19]. 2D materials, such as transition metal dichalcogenides, boron nitride, and phosphorene possess some intriguing physical and chemical properties, rendering them as promising candidates for future electronic and optoelectronic applications[2,4,20-22]. Like bulk semiconductors, doping is a key process in these 2D materials for their device applications. Therefore, it is quite natural to extend the defect calculation approaches from 3D bulk materials to these 2D materials to gain fundamental understanding of their doping and defect properties. However, a direct employment of the bulk defect calculation methods to 2D or other low dimensional systems (quantum dots, wires, etc.) encounters a serious problem manifested as the divergence of formation energies of the charged defects, which can be understood as follows. Adopting supercell approach under periodic boundary conditions in all three dimensions, an unavoidable "vacuum" region is added in the DFT calculations to separate the 2D material from its periodic images. Such "vacuum" region leads to a remarkable dissimilarity between a real charge put on host band-edge states and a virtual jellium charge. Specifically, when the jellium model is used, a virtual charge is uniformly filled in the whole supercell, including the vacuum region. This leads to a divergent Coulomb interaction between the jellium charge and the charge left on the 2D slab[17]. In this case, the jellium model also becomes unphysical because the jellium is very different from the real charge distribution when electrons are excited to the band edge states with their charge distribution confined within or near the 2D slab (Fig.1c). Therefore, unlike for the 3D case, for the low dimensional system, the jellium charge is not a good approximation of the real band edge states in a supercell calculation. Over the years, some models have been proposed to overcome the above discussed problems, including artificially constraining the background charge into a given region[23-27], or carrying out a posteriori correction[28-36]. However, all of these charged defect calculation methods for the low-dimensional systems are of some conceptual issues and computationally difficult to converge[37]. So far, there is no straightforward and rigorous theory to calculate the charged defects of low-dimensional structures. The lack of a universal method for both 3D and low-dimensional semiconducting materials has hindered the progress of first-principles defect studies.

In this paper, we develop a physical and unified approach to calculate the formation energy and transition energy levels of charged defects in both 3D bulk and low-dimensional semiconductors. For the 3D semiconducting materials, we find this method reproduces the same accuracy as the current widely-used jellium model. However, for the low-dimensional structures, it remarkably avoids the divergence induced by the artificial long-range Coulomb energy of the jellium model. Specifically, in our method, the ionized electron is represented by the real *host* CBM or valence band maximum (VBM) state (it can be easily extended to other states which have a statistically average energy $\varepsilon_F$ at finite temperature). Unlike in some of the previous pseudopotential calculations, in which the added jellium charge is treat artificially, thus, it is used only as compensating charge to prevent the divergence of electrostatic energy but not included in the calculation of exchange-correlation energy[38,39], in our case, the ionized electrons is treated in the same footing as other occupied electronic states and is included in all the total energy calculations. This is justified by noticing that if the removed electron is added back to the same states where it is removed, no change in total energy should occur. Our method is applicable for charged defect calculations of bulk and low-dimensional systems, including quantum dots, nanowires, surfaces and interfaces, and 2D materials, and the convergence of the calculated results with respect to the supercell size is similar in all dimensions.

## 2 Results

**Real state model by transferring charge from defect to real host band edge states**

Our approach is based on the following concept of defect ionization that is applicable for both bulk or low-dimensional semiconductors, that is, during the defect ionization, the carriers from the defect states are excited to the unperturbed host band edge states in the limit of infinite supercell size. Inspired by this concept, we represent the charge density of the ionized carrier by unperturbed host band edge state, e.g., CBM or VBM state, and treat it in the same footing as all the other occupied states in the self-consistent total energy calculation, thus realize the real process of carriers exciting to host band edge states from the defect level in the same supercell (denoted as transfer to real state model or TRSM for convenience). Apparently, in this treatment, the whole system is charge neutral because the charged defect and the excited carrier(s) are kept in the same supercell. To implement this TRSM process, we will revise the standard formula given in previous literatures for defect formation energy[17,37], that is, for a defect $\alpha$ in a charge state $q$, the formation energy is described as

$$\Delta H_f(q,\alpha) = \Delta H_f\left(q_{\varepsilon_D \to \varepsilon_{C,V}^h},\alpha\right) + q(\varepsilon_F - \varepsilon_{C,V}^h). \tag{1}$$

Here,

$$\Delta H_f\left(q_{\varepsilon_D \to \varepsilon_{C,V}^h},\alpha\right) = E_{\text{tot}}(q_{\varepsilon_D \to \varepsilon_{C,V}},\alpha) - E_{\text{tot}}(\text{host}) + \sum_i n_i(E_i + \mu_i), \tag{2}$$

where $q$ is the number of electrons taken from the defect state $\varepsilon_D$ and placed on the host CBM ($q > 0$) or VBM ($q < 0$) in the same defect supercell. $E_{\text{tot}}(q_{\varepsilon_D \to \varepsilon_{C,V}},\alpha)$ is the total energy of the supercell containing dopant $\alpha$ in which $q$ electrons are taken from the defect state $\varepsilon_D$ and placed on the host CBM $\varepsilon_C^h$ or VBM $\varepsilon_V^h$ state. $E_{\text{tot}}(\text{host})$ is the total energy of a supercell for perfect host crystal. $\mu_i$ is the chemical potential of constituent $i$ referenced to elemental solid/gas with energy $E_i$. $n_i$ is the number of elements removed from the host in creating the defect $\alpha$. The eigenvalues between different cells should be aligned with respect to same reference level. More detailed description of the calculation methods can be found later in the Method Section.

**Real state model for the three-dimensional charged defect calculations**

To check the validity of the jellium charge model with respect to the TRSM, we first compare the calculated formation energies and transition energies of defects in some prototype 3D semiconductors Si, BN, ZnO, and MoS$_2$ using both models. Figure 2 shows our calculated results using the standard jellium background charge and real host CBM or VBM charge (i.e. the TRSM), respectively. We see that the results calculated utilizing both approaches are in excellent agreement for various defects in 3D materials with the difference less than a few meV. Such good agreement is due to the fact that for sufficiently large supercells, the added jellium state is a good approximation for the host delocalized states like the conduction and valence band edge states in conventional semiconductors. It's worth noting that, in our TRSM model, the ionized electrons represented by the unperturbed host CBM or VBM states are treated in the same footing as other occupied electronic states and are included in all the total energy calculations, and thus it is more realistic and accurate. It is also worth noting that in the TRSM, the ionized charge is represented by the unperturbed host VBM or CBM states, not the lowest unoccupied state or highest occupied state in a defect supercell as carried out in some previous calculations[37], which can introduce large error if the cell size is not extremely large[16].

**Real state model for the low-dimensional charged defect calculations**

Next, we apply both models to 2D materials. Figure 3 shows the formation energies calculated by jellium model and our TRSM for the $C_B^{1+}$ and $C_N^{1-}$ charged defects in the 2D monolayer BN. It is clear that for both donor $C_B^{1+}$ or acceptor $C_N^{1-}$, the formation energies based on jellium model diverge as the vacuum layer thickness L$_z$ increases (Figs. 3a and 3b) because the poor screening of the JM introduces the artificial long range electrostatic interaction between the periodic images and the jellium compensating background charge that non-physically extends into the vacuum region[17]. However, in our TRSM, it has an excellent convergence with respect to L$_z$ when the lateral area $S$ (i.e., $L_x \times L_y$) is sufficiently large. For example, for the $C_N^{1-}$, the calculated formation energy has almost converged when $L_z$=10 Å. This means the electrostatic interaction error of the JM is eliminated in our more realistic TRSM approach. This is because in the TRSM, the real host CBM or VBM state is occupied. Unlike the nonphysical jellium state, it is localized in the 2D BN itself (Fig.1c) and independent of the $L_z$, thus, it increases the screening in the $L_z$ direction, and meanwhile eliminates the divergence of the electrostatic energy caused by uniform jellium background charge in the whole supercell in the jellium model. For the in-plane region, because the ionized charges are located in the 2D slab, its convergence is like the case in the 3D bulk system, so the in-plane convergence of the TRSM for the 2D systems is also excellent. Consequently, we confirm that the TRSM model is more physical and reliable. Figure 3c shows that using the TRSM, the calculated (+/0) transition energy of donor $C_B$ and (0/-) transition energy of acceptor $C_N$ in monolayer BN is 1.72 eV below the CBM and 1.60 eV above the VBM, respectively. This is in reasonably good agreement with the previous calculation in which the donor $C_B$ (+/0) and acceptor $C_N$ (0/-) transition is at about 2.0 eV below the CBM and 1.8 eV above the VBM[17]. The difference can be understood as follows: in Ref. 17, the unphysical JM is still used and the results are conditionally converged which still underestimate the Coulomb interaction between defect charge and ionized charge. In our TRSM method, the more physical and realistic ionized charge is used, which lowers the energy of the charged states, so the transition energy levels are more shallow compares to the previous JM calculations. This indicates that in 2D charged state calculations, the JM should be fully avoided.

# 3 Conclusions

In summary, we developed a realistic dimension-independent approach for charged defect calculations in semiconductors, that is especially useful for low-dimensional materials such as quantum dots, nanowires, surfaces and interfaces, which suffers both conceptually and computationally in previous jellium model calculations. For the 3D semiconductor systems, this method produces similar results as the current widely-used jellium model, but for the low-dimensional structures, it is able to eliminate the divergence caused by the artificial electrostatic energy faced in the jellium model, and has an excellent convergence for the formation energy and transition energy calculations of charged defects. Our method can be applied to charged defect calculations for all low-dimensional systems and can be easily extended to including more occupied states and calculate the exciton binding energy if the supercell size is comparable to the exciton radius.

**Computational methods**

Our calculations are performed within first-principles density function theory (DFT) as implemented in *quantum-espresso* package[40,41]. The Perdew-Burke-Ernzerhof (PBE) functional[42] is used for exchange and correlation potential, and only the Γ point for the Brillouin zone integration. The size of supercell is chosen to ensure the results are converged unless mentioned otherwise. The norm-conserving pseudopotentials[43,44] for treating the valence electrons are used. The kinetic energy cut off for the plane wave basis set is 65 Rydberg, and the total energy threshold for convergence is $10^{-12}$ Rydberg. All atoms are fully relaxed until the Hellman-Feynman forces acting on each atom are less than 10⁻¹⁰ Rydberg/Bohr. Figure 4(a) and (b) show the self-consistent calculation flow for donor and acceptor charged defect total-energy calculations, respectively, in TRSM. $\psi_D$, $\psi_{CBM}^{host}$, and $\psi_{VBM}^{host}$ are the wave functions of defect state, host CBM and host VBM, respectively. $q$ is the number of electrons that are excited from the defect state $E_D$ to the host CBM for donor ($q > 0$) or from the host VBM to the defect state $E_D$ for acceptor ($q < 0$). $N$ is the total number of electrons in the system. In practice, we first calculate the host CBM (or VBM) charge distributions $|\psi_{CBM}^{host}|^2$ (or $|\psi_{VBM}^{host}|^2$) in pure host systems. Then, for the total energy calculations in the charged defect systems, we remove (add) the defect charge distribution $|\psi_D|^2$ with integral charge $q$ and add (remove) the fixed host CBM (VBM) charge distributions $\psi_{CBM}^{host}$ ($\psi_{VBM}^{host}$) with same total charge $q$ in each self-consistent step until the total energy converges. Therefore, in the TRSM model, the ionized electrons are treated in the same footing as other occupied electronic states in the total energy calculations.

**Acknowledgements**

This work was supported by the Science Challenge Project (Grant No. TZ20160003), the National Key Research and Development Program of China (Grant No. 2016YFB0700700), the National Natural Science Foundation of China (Grants No. 61922077, No. 11704114, No. 11874347, No. 11804333, No. 61121491, No. 61427901, No. 11474273, and No. U1930402), China Postdoctoral Science Foundation Funded Project (Grant No. 2017M620872). H.-X. D. was also supported by the Youth Innovation Promotion Association of Chinese Academy of Sciences under Grant No. 2017154.

**Author contributions**

J.X. and K.Y. performed defect formation energy calculations and prepared the figures. J.X., K.Y., D.G., and T.S. revised DFT code to perform TRSM calculations. H.-X.D., S.-H.W., and J.-W.L. proposed the research project and guided in detail for progress. H.-X.D. S.-H.W., S.-S.L., and J.-W.L. conducted the results analysis, discussion, and writing of the paper.

**Additional information**

Competing financial interests: The authors declare no competing financial interests.

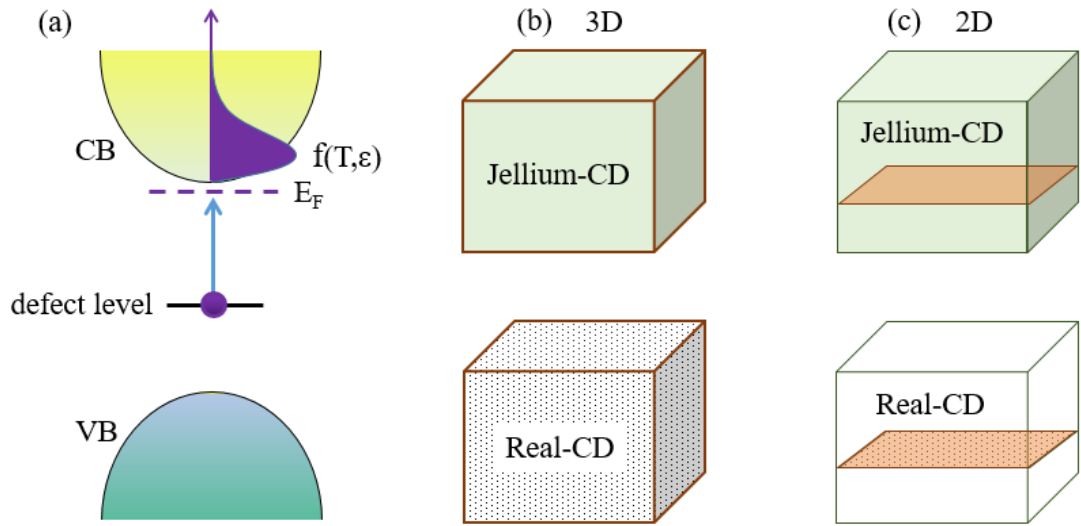

**Figure 1.** (**a**) Schematic plot of occupations of the ionized electronic states above the CBM with a statistical distribution for a given $E_F$ at the finite temperature from the donor level. (**b**) and (**c**) Schematic plots of charge distributions of jellium charge distribution (Jellium-CD) and real state with a certain statistical charge distributions (Real-CD) in the 3D and 2D semiconductors, respectively.

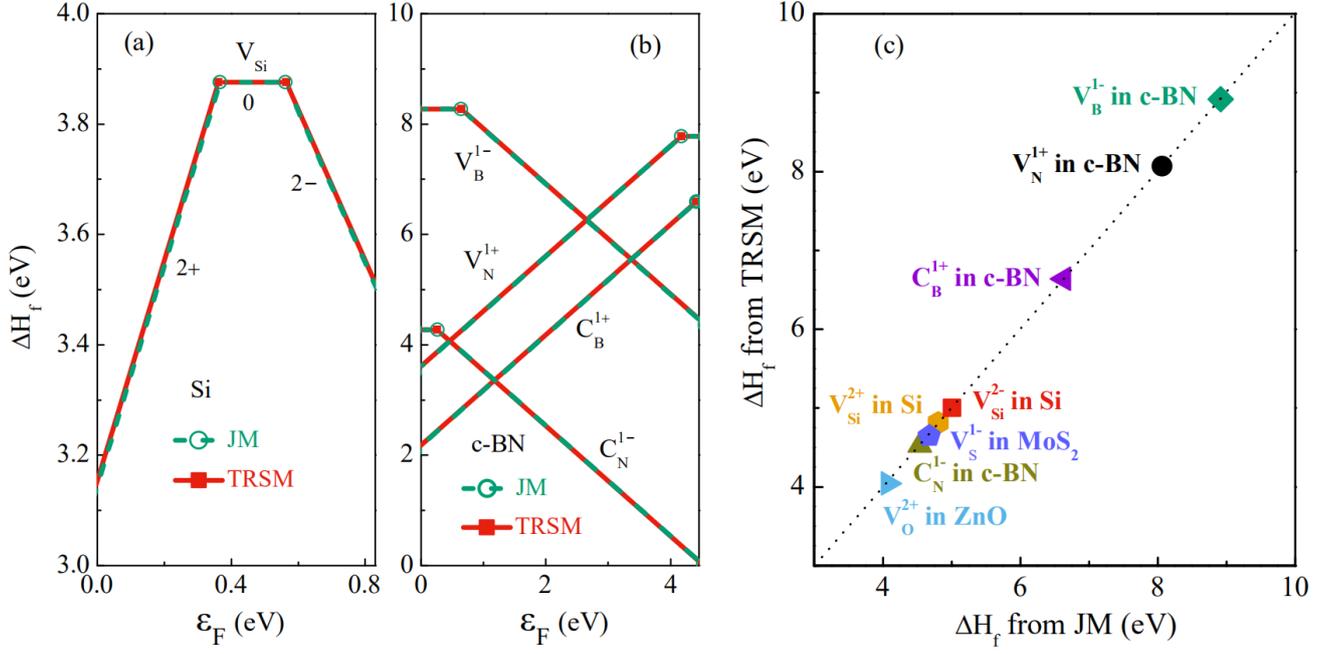

**Figure 2:** Comparisons of the formation energies of charged defect as a function of Fermi energy $\varepsilon_F$ in 3D system calculated by jellium model (JM) and TRSM model. (**a**) The formation energies of Si vacancy ($V_{Si}$) in bulk Si. (**b**) N and B vacancy ($V_N$ and $V_B$) and C atom substituting at B site ($C_B$) or at N site ($C_N$) in cubic BN (c-BN). (**c**) Comparisons of the calculated formation energies of charged defects using the JM and TRSM in bulk Si, BN, ZnO, and MoS2. Here all the calculations are converged with respect to the supercell size and in the conditions $\mu_i = 0$ and $\varepsilon_F = \varepsilon_{CBM}$ or $\varepsilon_{VBM}$ for $q > 0$ or $q < 0$ defects, respectively.

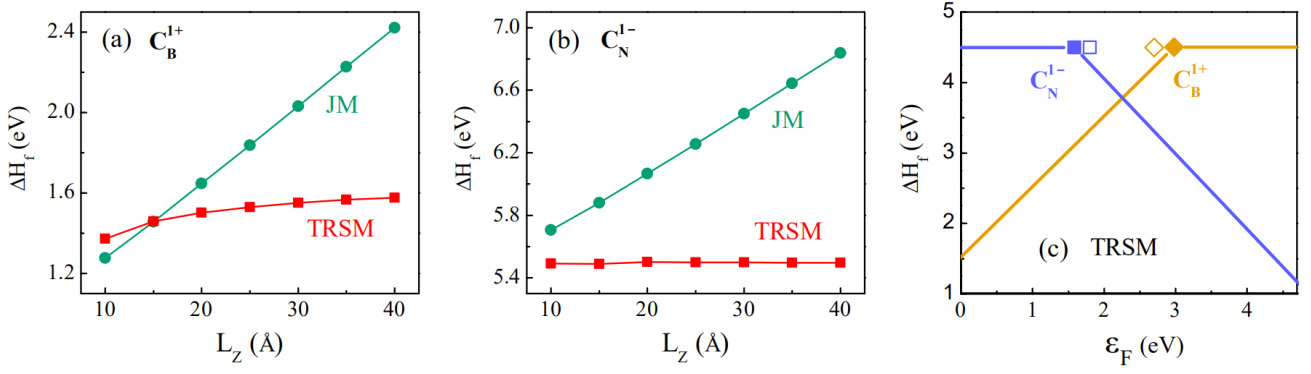

**Figure 3:** Formation energies of charged defect in 2D system calculated by jellium model (JM) and our TRSM model. (**a**) and (**b**) are the formation energies of $C_B^{1+}$ and $C_N^{1-}$ in monolayer BN sheet as a function of layer-layer separation $L_z$ with large lateral supercell area S (9×9) and $\varepsilon_F = \varepsilon_{VBM}$, respectively. (**c**) The calculated formation energies and transition energy levels of $C_B$ and $C_N$ in monolayer BN using our TRSM model as a function of Fermi energy $\varepsilon_F$. As a compassion, we also show the (+/0) transition energy of acceptor for $C_B$ (empty diamond dot) and the (0/-) transition energy of acceptor for $C_N$ (empty square dot) in monolayer BN calculated by D. Wang *et al.*[17]. Here all the calculations are in the conditions $\mu_i = 0$.

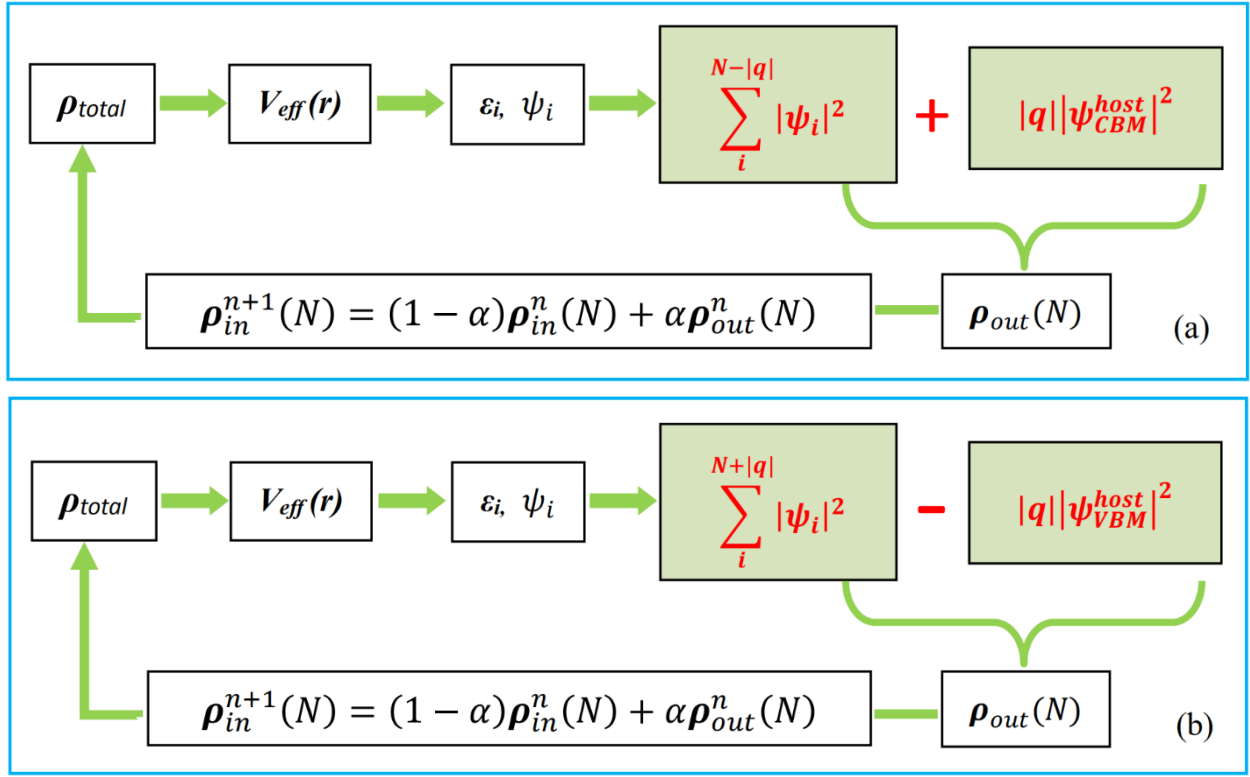

**Figure 4.** The sketch of self-consistent calculation flow for (**a**) donor and (**b**) acceptor charged defect total-energy calculations, respectively, in TRSM approach.